# A New Secure Mobile Cloud Architecture

Olayinka Olafare[1*], Hani Parhizkar[1] and Silas Vem[1]

[1] School of Computer Science, University of Nottingham Malaysia Campus, Semenyih, Selangor 43500, Malaysia.

**Abstract**

The demand and use of mobile phones, PDAs and smart phones are constantly on the rise as such, manufacturers of these devices are improving the technology and usability of these devices constantly. Due to the handy shape and size these devices come in, their processing capabilities and functionalities, they are preferred by many over the conventional desktop or laptop computers. Mobile devices are being used today to perform most tasks that a desktop or laptop computer could be used for. On this premise, mobile devices are also used to connect to the resources of cloud computing hence, mobile cloud computing (MCC). The seemingly ubiquitous and pervasive nature of most mobile devices has made it acceptable and adequate to match the ubiquitous and pervasive nature of cloud computing. Mobile cloud computing is said to have increased the challenges known to cloud computing due to the security loop holes that most mobile devices have.

## 1. Introduction

A paradigm shift in the computing and Information Technology world is being envisaged, but this is slow paced primarily due to the challenges imminent to this emerging technology, despite the fact that it also carries along with it immense benefits that outweighs those which have been used or experienced in the past (Zissis & Lekkas, 2012). Stakeholders and interest groups hold differing opinions as to the level of importance and impact this new technology will have on business processes and activities. This new archetype is generally referred to as "Cloud Computing". Cloud computing is said to be the way forward and the solution with huge profitability hence resource sharing/pooling, on-demand self-service, broad network access, rapid elasticity/scalability are taken into consideration (Subashini & Kavitha, 2011). These are some well-established benefits of cloud computing but it is worth mentioning that the major concern of stakeholders in this niche given the mouthwatering benefits earlier outlined is that of "*security*" (Mather & Kumaraswamy, 2009; Verma & Kaushal, 2011). The service provision and support of this new technology extends to both mundane computers and enthusiastic mobile computers like smart phones, iPads and tablets (Horrow, et al. 2012). Based on its support for desktop computers and mobile computers, a two-phased security challenge is associated with the cloud computing paradigm, ***Mobile Cloud Computing*** (MCC) and ***Cloud Computing*** (CC) as a technology. This research is focused on the security challenges and possible solutions in these two areas. The cloud computing model extends its enormous potentials/benefits to mobile device users, hence the development of what is termed an "offspring technology"; MCC riding on the potentials offered by the cloud computing model. MCC is defined as a new technology that provides cloud computing resources and services to mobile device users (Popa, et al. 2013). Just like cloud computing, this new technology has different sides and definitions to it. Another author defines mobile cloud computing as: *„an infrastructure where both the data storage and data processing happen outside of the mobile device. Mobile cloud applications move the computing power and data storage away from mobile phones and into the cloud, bringing applications and MC to not just Smartphone users but a much broader range of mobile subscribers"* (Dinh, et al. 2011). From a careful review of several related literatures we define mobile cloud computing (MCC) as "*a robust and flexible mobile computing technology that leverages the vast potentials and elastic resources of the cloud computing and network technologies toward unlimited functionalities, storage, and mobility to serve multiple mobile devices anywhere, anytime using the internet as a backbone regardless of heterogeneous environments and platforms*". A pictorial definition is shown in Figure 1 below;

*Corresponding author

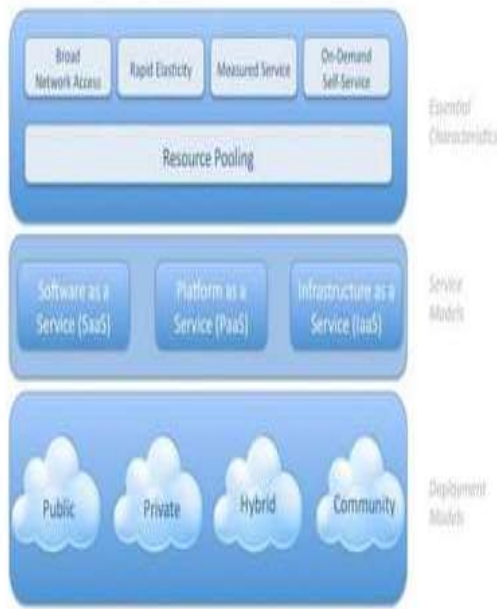

Figure 1 NIST Visual Model of Cloud Computing Definition

## 2. Background of Mobile Cloud Computing

MCC is a relatively new technology leveraging on the bounty benefits of an older ubiquitous technology; Cloud Computing. MCC is a technology which has come as a solution to the challenges faced by mobile devices; smartphones, PDAs, tablets etc., which include limited computing power and inadequate storage capacity amongst many others (Popa, et al. 2013). *MCC refers to an "infrastructure where data processing and data storage is done outside the mobile device. In MCC, computing power and data storage is executed away from mobile phones and into the cloud, MCC is not just for Smartphone users but also covers a wide range of mobile subscribers"* (Prasad & Gyani, 2012). The advent of MCC (Popa et al., 2013) is seen as a solution in the right direction to the afore- mentioned challenges faced by mobile device, as it takes advantage of Cloud Computing resources, thereby providing mobile device users with the power and unlimited resources of the cloud. There is no mobile cloud computing without cloud computing as such we will be making reference to both technologies in the course of this review. Mobile cloud computing further extends the benefits of cloud computing, owing to the improved features offered on mobile devices like smartphones, PDAs and tablets. Studies show that there has been a surge in the use of smart-phones and that this has continued to grow over the years. A study made by Gartner (Gartner, Inc., 2011) shows that in the third quarter of 2011 the sale of smart-phones increased by 42% worldwide. To further support this assertion, a leading research group Allied Business Intelligence (ABI) says (ABI Research, 2013), by 2015 more than 240 million business customers will use Cloud resources and services through mobile devices and this will result in revenues of billions of dollars. The potential benefits of mobile cloud computing is hampered by the perceived security challenges imminent and peculiar to a distributed, pervasive, parallel and ubiquitous computing paradigm, known as cloud computing, which directly shares its burden with MCC (Kim et al. 2013; Popa, et al. 2013).

In view of this is our research work is set to propose a secure mobile cloud architecture which will enable the successful and secure delivery of the potential benefits of MCC to an eager and expectant consumer. Figure 2 shows how mobile devices can be used to access the immense benefits of the cloud environment and resources. The advent of MCC came into center stage shortly after the evolution of Cloud Computing (CC) which is barely a decade old. Upon the realization of the potential benefits of Cloud Computing and the desire to further extend these benefits from fixed computer systems to mobile devices brought to limelight MCC, (Dinh, et al. 2011); (Wang, et al. 2013) which has been tipped to bridge the gap that exists in using desktop computers to access CC resources. In February 2012 a press release made by a smart market insight group Canalys reported that (Canalys, 2012) worldwide shipments of smartphones amounted to (487.7 million), exceeding PCs (414.6 million including tablets) in 2011. A recent research conducted by leading research organization Juniper Research has predicted that (Juniper, 2011) revenue from mobile enterprise cloud-based applications and services is expected to rise from nearly $2.6 billion in 2011 to $39 billion in 2016. (Markets And Markets, 2010) MarketsAndMarkets.com, a global technology forecast company in August 2010 published on their website a forecast of the global mobile applications market saying it is expected to be worth $25.0 billion by 2015. Owing to these forces, there is a desire to be able to use mobile devices to access the vast potentials of the cloud, as the (Wang, et al. 2013) use of mobile cloud computing will enable more powerful applications, and hence more significant growth.

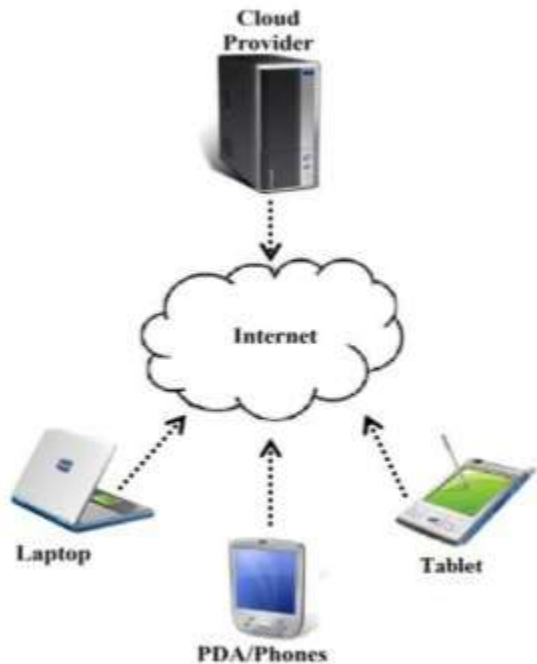

Figure 2 Mobile Cloud Computing (Gharehchopogh, Rezaei, & Maleki, 2013)

### 3. Applications of Mobile Cloud Computing

The need and importance of mobile cloud computing cannot be over-emphasized considering its wide application and its unique empowered advantages (Fernando, et al. 2013). The application of mobile cloud computing cuts across most of our daily human activities and in most cases comes in very handy to meet the need of its users in providing flexibility, information on the go; anywhere, anytime with very little financial implication, considering that cloud services are priced on a pay per usage scheme. Its application as seen in many literatures is briefly described below:

#### 3.1 Image Processing

Graphical Optical Character Recognition (GOCR) program was (Chang, 2005) conducted on mobile devices by a group of researchers. Its real life application can be sighted by an instance where a tourist in a foreign country takes pictures of road signs and performs Optical Character Recognition (OCR) to translate the words to a language he/she understands. A mobile device based on its potentials and resources might not have adequate data processing capacity to successfully perform the OCR, but given access to a mobile cloud, this task can be easily and quickly achieved.

#### 3.2 Mobile Healthcare (M-health)

Health they say is wealth (research2guidance, 2013) so considering how important it is to stay healthy and wealthy, having access to these basic human needs has never been more important considering the various health challenges people are exposed to due to the high level of industrialization as seen in our world today. M-health applications involves (but is not limited to) the use of mobile devices in collating community and clinical health data, delivery of healthcare information to professionals, researchers and patients, real-time monitoring of patient vital signs, and direct provision of care. (Kohn, et al. 2003) (Figure 3). The application of MCC comes in handy to mitigate the limitations of conventional medical treatment like storage capacity, *security and privacy.*

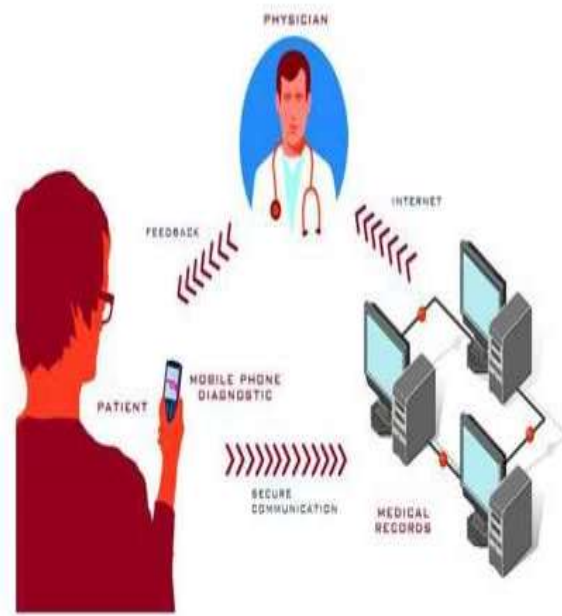

Figure 3 Mobile Healthcare (Applied Nanodetectors, 2013)

#### 3.3 Mobile Learning

Mobile learning (Figure 4) also written as m-learning is an off-shoot of electronic learning (elearning) exploiting the advantages of mobile devices like compactness and mobility. In view of the limitations of conventional m-learning technology, like increased cost of network devices, slow data transmission rate (Chen, et al. 2010; Li, 2010), we are presented with Mobile Cloud Learning (MCL) as a solution to the challenges that face orthodox m-learning technology (Yin & David, 2009). This improvement provides learners with better services like information processing speed and longer battery life.

## 4. Security of Mobile Cloud Computing

Mobile cloud computing consist of two major components; (1) the mobile device, (2) the cloud. The latter component and some aspects of the former have been discussed in the preceding sections. In our review it has been mentioned that to achieve a secure mobile cloud architecture, it is imminent that the cloud is secure, in other words, secure mobile cloud cannot be achieved without first ensuring a secure cloud environment. On this premise, have we discussed extensively measures needed to ensure the security of the cloud. This section discusses the security issues/challenges of the *mobile* cloud which we believe combined with the knowledge gathered from the reviews carried out on the cloud security, will ensure success in designing a *secure mobile cloud architecture.* Basically the security issues in mobile cloud computing is associated with (1) security issues in the cloud, (2) security of the mobile device and (3) the security of the communication channel between the cloud resources and the mobile device (Popa, et al. 2013).

### 4.1 Security of Mobile Device

There is an extent to the amount of security that can be offered to a mobile device user to enhance his/her experience in the cloud environment, so we are limiting our solution and exploration to security concerns that are within our reach. For instance physical attacks like theft on a mobile device are out of the context of this research work. Mobile devices have a number of applications running on them; some of which are in-built with the mobile device operating system, while some are third party applications developed by various vendors. Our concern is more on the third party applications *user service agreement* that is how much information should the third party application have access to on the mobile device (Prasad & Gyani, 2012). We propose the adoption of an application on the mobile device with the potential of checking and restricting the amount of information third party applications can have access to. So we call this a *user service agreement validating application* (Modi, et al. 2013)*.* Besides this, we also need to check the validity and authenticity of third party application before they are installed on the mobile device. Malicious codes and application programs also have to be checked so we propose an *application integrity* checking program which will be used to check for similarities in third party applications when they are being updated. Our proposed program/application in processing and updating request for a third party application will check the third party application signature or certificate to ensure that the updated version's signature matches the original signature of the third party application that was installed initially. Updating a third party application will be blocked if the signature does not correspond to the signature of the earlier version of the application. So for mobile device security we propose these;
(This can be achieved with ease considering that our proposed solution to cloud security provides an option for categorization of data into: not sensitive, sensitive and highly sensitive).

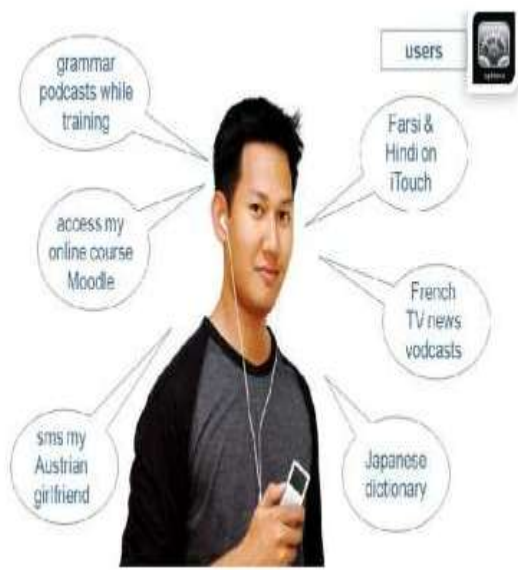

Figure 4 Mobile Learning (Hockly, 2009)

2013).

### 4.2 Security of Communication Channel

When the cloud and the mobile device are secure, one might be tempted to assume all is well, but that is not totally correct, because a lot of vulnerabilities may still occur during data transfer from the cloud to the mobile device and vice versa. As such for us to have a higher level of security we propose the security of the communication channel. When the mobile device makes contact with the cloud resources, request is sent and the cloud responds by sending back the information or data requested by the mobile device. In this process, it is easy for an attacker to bridge the data transmission and act as a cloud server, so the user ends up losing his/her data. To mitigate this anomaly, we propose the use of SSL (Secure Socket Layer) certification and this is part of our architecture layout. To ensure this, network security Certificate Authorities (CAs) issue organizations with unique certificates, which validate the existence of such organizations in the online community as trusted organizations. Each organization is issued a unique domain as a ticket for transaction. When a potential user signs up with an organization with a unique domain for cloud services, the cloud service provider sends a unique identification information and its SSL certification details to the potential user, which can be verified by contacting the relevant agencies. Once the user starts using the services of the cloud service provider, data sent to the user will be an SSL encrypted data or information with the key for decrypting the information is sent to the user's personal email account. By adopting this measure for securing the communication channel, we strongly believe that the security of mobile cloud computing will be enhanced (Sood, 2012). All of the security measures discussed will be implemented or taken into consideration in the design of our proposed architecture. In the next section we have discussed some existing architectures proposed in some of the articles we reviewed in the process of conducting this research.

## 5. Mobile Cloud Computing Models/Architectures

In the preceding sections, we have listed and briefly described some of the benefits and applications of MCC all of which amounts to nothing if a technology waits to jeopardize the resources and information of potential users if the activities and transactions in MCC cannot be done *securely.* Of what use is it to have a pool of funds without adequate security or a job without security? Some would say the rather not have such. According to (Alzain, et al. 2012)**,** (Gonzalez, et al. 2011) (Khan, et al. 2013) **security** is said to be one of the most critical areas of concern in Cloud Computing (CC). Based on this concern and the desire to deliver the afore-mentioned benefits and applications of MCC to a technology driven society, this research work proposes *a secure mobile cloud architecture.* At present there are several Mobile Cloud Architectures (MCA)/Mobile Cloud Models (MCM). In carrying out this research, we conducted a series of tests and analyses relying on expert opinion using the Delphi technique to decide and propose a *New Secure Mobile Cloud Architecture*. Listed below are some existing mobile cloud architectures:

5.1 Mobile Cloud Computing (General) Architecture.

A general architecture for mobile cloud computing was proposed by (Prasad & Gyani, 2012) (Dinh, et al. 2011). This model is designed following these communication channels; mobile devices are connected to the mobile networks via base stations (e.g. base transceiver station (BTS), access point, or satellite) as shown in the diagram, which is responsible for establishing a connection between the mobile device(s) and creates a functional interface between the network and the mobile device(s) and also controls the connections. When a mobile user sends a request, information about the particular mobile user (e.g. ID and location) is transmitted to the central processor which is connected to a server which provides mobile network services. Through this connection medium, mobile network operators can offer AAA (Authentication, Authorization, and Accounting) services to mobile users, based on the home agent (HA) and subscriber's data stored in the database. After this connection and information exchange phase, the subscriber's requests are delivered to the cloud through the internet. In the cloud, controllers process the requests to provide mobile users with the corresponding cloud services. These services are developed with the concepts of utility computing, virtualization and service oriented architecture (e.g. web application and database servers). The general architecture is the basis for all redefined architectures, which is a product of various researches to ensure security and a more improved mobile cloud architecture considering that the general architecture has unresolved security concerns which leave potential users vulnerable to attacks. *The proposed "Mobile Cloud Computing General Architecture" has very little provision for user data security in the cloud or on the mobile device. This leaves a lot of room for vulnerability on the part of the users, which is why most users are yet to conform to the services of mobile cloud.*

5.2 Secure Mobile Cloud (SMC) Architecture

The Secure Mobile Cloud Framework is a component based model, where each component plays its role at different locations and has a particular functionality or security concern it addresses. This model looks into security issues on both the cloud platform as well as the mobile device. On the mobile device, the model has the following;

I. Optimization manager:
II.     Mobile manager
III.    Policy manager
IV.    Mobile security manager

And on the cloud side the model comprises of the following components;

I. Application manager
II.    Policy manager
III.    Cloud security manager (Popa, et al. 2013)

The proposed framework/model is designed to ensure the integrity of an application setup and to secure the communication between the same application components (i.e. between components running on the mobile side and those running in the cloud, and among the components running only in cloud). *This framework does not ensure security of data transmitted between components on the mobile device and it also does not provide secure data transmission between cloud components.*

5.3 Cloud Computing Secure Architecture on Mobile Internet

The architecture proposed by (Xiu-feng, 2011) is partitioned into three major layers; the first layer comprises of cloud secure application service resources which includes *privacy data protection, cipher-text data query, data integrity validation, security event early warning and content security service.* The Second layer consist of a group of cloud secure platform service resources for cloud computing virtualization, which host the following; *virtual machine segregation, virtual machine secure monitoring, virtual machine secure migration and virtual machine secure mirror.* The third layer is where the infrastructure is based, and there is little or no security measure in place to ensure the security of the infrastructure *which is one of the areas our proposed architecture will provide security for*. Besides the three major layers, there is a partition for security management with security provisions like firewall, user management, key management, early warning etc. From our review, we noticed that there is not a single architecture that provides solution for all the above listed security issues, possibly because deploying all of this security protocols to mitigate the security issues in the cloud as well as mobile cloud might not be feasible or cost effective as such, each security architecture addressed different security concerns. In our proposed architecture, we designed questionnaires for particular target respondents, "experts" and used their opinions to decide the impact factor or risk rating of the above listed security concerns using the Delphi survey technique.

## 6. Identification of Project Phases

In this section, we have listed the major phases of this project.

**I. Fact finding phase:** This was done majorly through the use/review of scholarly articles relevant to the chosen research topic, we also engaged lecturers and students of the University of Nottingham Malaysia Campus (UNMC) in an informal chit-chat session to find out their perception/opinion on issues bothering both cloud computing and mobile cloud computing as well as their perceived importance and impact of this new technology on everyday life of an average person.

**II. Requirement analysis phase:** This phase involved gathering information on the basic requirements (security concerns or issues) for ensuring safety in the cloud, thereby providing us with relevant information/literature on security issues that need be addressed and the weight or impact factor of such security concerns. A questionnaire based on these requirements was then designed and sent a group of fifteen experts of mobile cloud computing to further confirm these requirements/security concerns were genuine. The shortlisted security concerns were verified as being genuine concerns by all the experts; after which the experts went a step further to give individual opinions on the impact factor of each of these security concerns on mobile cloud computing (Table 1).

**III. Design phase:** The design phase is the stage where the information gathered from the fact finding and requirement phases of the project was being implemented to meet the specifications gathered from end users, service providers and experts. Based on the opinion of the group of experts and reviewed articles, we designed the proposed secure mobile cloud architecture ensuring it contains adequate security components to mitigate the perceived security concerns/issues as raised by the interest groups.

## 7. SPSS (Statistical Package for Social Sciences) Analysis of the Questionnaires

7.1 Round One Questionnaire Analysis

The importance or impact factor of the listed security components is valued on a scale of 1 to 5 Maximum (max) and Minimum (min) values as assigned to each security component by the respondents.

The sum total of each component as well as the weighted average (mean), which also represents the content value of each security component.

Table 2 compares the content value (mean) of each security component; this information will be used in deciding the components that will be of greater priority in the design of the secure mobile cloud architecture. Figure 5 is a graphical representation of the content value of each of the security components to aid in better understanding the impact factor of the security components. These results will be compared with those of the second round questionnaire in subsequent sections to check for disparity or variance in the opinions of the respondents. The outcome of this comparison is what determined if there would be need for further rounds of questionnaires.

Table 1 Descriptive Statistics Summary of Security Components.

| Security Component | N | Min | Max | Sum | Mean |
|---|---|---|---|---|---|
| Authentication | 15 | 4 | 5 | 72 | 4.80 |
| Authorization | 15 | 3 | 5 | 69 | 4.60 |
| Repudiation | 15 | 3 | 5 | 64 | 4.27 |
| Network Security | 15 | 4 | 5 | 73 | 4.87 |
| Data Security | 15 | 3 | 5 | 68 | 4.53 |
| Application Integrity | 15 | 2 | 5 | 69 | 4.60 |
| User Management | 15 | 2 | 5 | 61 | 4.07 |
| Data Encryption | 15 | 3 | 5 | 64 | 4.27 |
| Intrusion Detection | 15 | 4 | 5 | 69 | 4.60 |
| Virtual Machine Segregation | 15 | 1 | 5 | 56 | 3.73 |
| Virtual Machine Monitoring | 15 | 2 | 5 | 56 | 3.73 |
| Virtual Machine Migration | 15 | 2 | 5 | 54 | 3.60 |
| Valid N (list wise) | 15 | | | | |

N represents the number of responses received

Table 2 Comparison of the Content Value (Mean) of the Security Components.

| Security Components | Mean | N |
|---|---|---|
| Authentication | 4.80 | 15 |
| Authorization | 4.60 | 15 |
| Repudiation | 4.27 | 15 |
| Network Security | 4.87 | 15 |
| Data Security | 4.53 | 15 |
| Application Integrity | 4.60 | 15 |
| User Management | 4.07 | 15 |
| Data Encryption | 4.27 | 15 |
| Intrusion Detection | 4.60 | 15 |
| Virtual Machine Segregation | 3.73 | 15 |
| Virtual Machine Monitoring | 3.73 | 15 |
| Virtual Machine Migration | 3.60 | 15 |

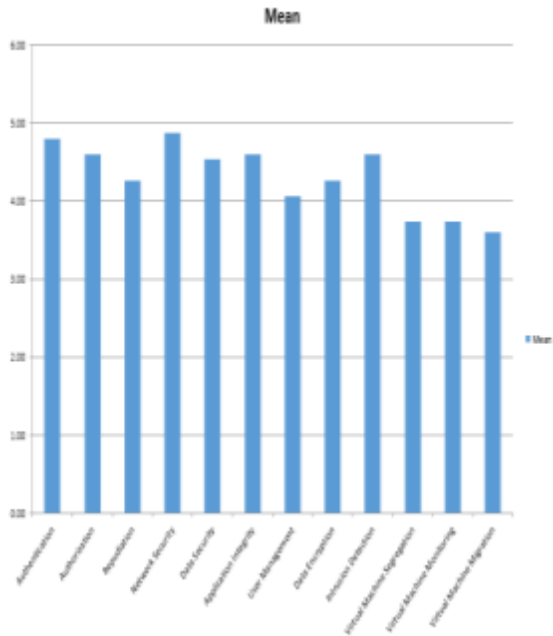

Figure 5 Histogram Comparing the Content Value (Mean) of the Security Components.

### 7.2 Reliability Analysis of the Questionnaire Response (Round One)

Reliability means a measure (in our case questionnaire) that should consistently reflect the construct that it is measuring. To check for the reliability of the questionnaire used, we subjected the questionnaire response to the Cronbach's alpha, α reliability test;
Cronbach's alpha, α reliability test; $\alpha = N^2 \overline{Cov} / \Sigma s^2 + \Sigma Cov$,
Where; N is the number of items, $\overline{Cov}$ is the average covariance, $S^2$ is square of item variance, Cov is the covariance.

A Cronbach's alpha value of 0.7 to 0.8 is an acceptable value which qualifies the data values used as reliable (Field, 2006). The Cronbach's alpha value was found to be 0.713 (for 12 items) which validates the reliability of the questionnaire response used in this project. So from the above statistical analysis/calculation, it can be seen that our data sets are very reliable. In Table 3, the values in the column labeled Cronbach's alpha if item deleted are the values of the overall "α" if that particular component (item) is not included. This shows a change in Cronbach's α that will be seen if a particular component is deleted. Similarly, the values in the subsequent columns show the effect of deleting any of the components on the scale mean and scale variance respectively.

Table 3 Security Components (Item) Total Statistics (Round Two)

| Security Component | Scale mean if item deleted | Scale variance if item deleted | Cronbach's alpha if item deleted |
|---|---|---|---|
| Authentication | 46.87 | 19.410 | .693 |
| Authorization | 47.07 | 20.210 | .722 |
| Repudiation | 47.40 | 20.114 | .725 |
| Network Security | 46.80 | 20.886 | .719 |
| Data Security | 47.13 | 20.267 | .731 |
| Application Integrity | 47.07 | 16.781 | .672 |
| User Management | 47.60 | 15.400 | .631 |
| Data Encryption | 47.40 | 21.400 | .755 |
| Intrusion Detection | 47.07 | 18.638 | .681 |
| Virtual Machine Segregation | 47.93 | 15.495 | .654 |
| Virtual Machine Monitoring | 47.93 | 15.067 | .641 |
| Virtual Machine Migration | 48.07 | 16.067 | .663 |

Table 4 Descriptive Statistics Summary of Security Components.

| Security Component | N | Min | Max | Sum | Mean |
|---|---|---|---|---|---|
| Authentication | 14 | 4 | 5 | 72 | 4.80 |
| Authorization | 14 | 3 | 5 | 69 | 4.60 |
| Repudiation | 14 | 3 | 5 | 64 | 4.27 |
| Network Security | 14 | 4 | 5 | 73 | 4.87 |
| Data Security | 14 | 3 | 5 | 68 | 4.53 |
| Application Integrity | 14 | 2 | 5 | 69 | 4.60 |
| User Management | 14 | 2 | 5 | 61 | 4.07 |
| Data Encryption | 14 | 3 | 5 | 64 | 4.27 |
| Intrusion Detection | 14 | 4 | 5 | 69 | 4.60 |
| Virtual Machine Segregation | 14 | 1 | 5 | 56 | 3.73 |
| Virtual Machine Monitoring | 14 | 2 | 5 | 56 | 3.73 |
| Virtual Machine Migration | 14 | 2 | 5 | 54 | 3.60 |
| Valid N (list wise) | 14 | | | | |

### 7.3 Round Two Questionnaire Analysis

Table 4 is a summary of the responses from the second round of questionnaire received from the expert respondents. Table 5 compares the content value (mean) of each security component for the second round of response received from the expert respondents. This information was used in deciding the components that will be of greater priority in the design of the secure mobile cloud architecture.

Figure 6 is a graphical representation of the content value of each of the security components for the second round of questionnaire response from the expert respondents for a better understanding of the impact factor of the security components. These results will be compared with those of the second round questionnaire in subsequent sections to check for disparity or variance in the opinion of the respondents. The outcome of this comparison is what determines if there will be need for further rounds of questionnaires.

Table 5 Comparison of the Content Value (Mean) of the Security Components

| Security Components | Mean | N |
|---|---|---|
| Authentication | 4.64 | 14 |
| Authorization | 4.50 | 14 |
| Repudiation | 4.14 | 14 |
| Network Security | 4.50 | 14 |
| Data Security | 4.71 | 14 |
| Application Integrity | 4.07 | 14 |
| User Management | 3.79 | 14 |
| Data Encryption | 4.43 | 14 |
| Intrusion Detection | 4.36 | 14 |
| Virtual Machine Segregation | 3.64 | 14 |
| Virtual Machine Monitoring | 3.86 | 14 |
| Virtual Machine Migration | 3.79 | 14 |

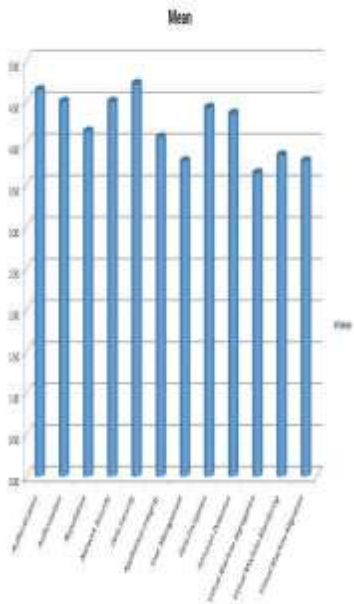

Figure 6 Histogram Comparing the Mean (Content Value) of the Security Components

### 7.4 Reliability Analysis of the Questionnaire Response (Round Two)

The Cronbach's alpha value for this round of questionnaire responses was found to be 0.869 (12 items) which validates the reliability of the round two questionnaire response used in this project. So from the above statistical analysis/calculation, it can be seen that our second round data sets are equally very reliable.

### 7.5 Checking for Consistency in the two Rounds of Questionnaire Response (Test-Retest Reliability)

In sections 4.3.2 and 4.3.4 we successfully validated the reliability of the data gathered from the two rounds of questionnaire responses. In this section, consistency or disparity in the two rounds of response was checked to determine if there was need for more rounds of questionnaire responses, taking into consideration the final data sets to be used in the design of the secure mobile cloud architecture. This means that all other factors being equal or kept constant, a respondent should get the same or similar score on a questionnaire if the questionnaire is completed at two different times, which is known as "test-retest reliability" (Field, 2009). Test retest reliability is the ability of a measure to produce consistent results when the same entities are tested at two different points in time.

Table 6 Security Components (Item) Total Statistics (Round Two)

| Security Component | Scale mean if item | Scale variance if item deleted | Cronbach's alpha if item deleted |
|---|---|---|---|
| Authentication | 45.79 | 44.181 | .865 |
| Authorization | 45.93 | 43.148 | .864 |
| Repudiation | 46.29 | 40.374 | .856 |
| Network Security | 45.93 | 42.841 | .858 |
| Data Security | 45.71 | 46.681 | .873 |
| Application Integrity | 46.36 | 39.324 | .852 |
| User Management | 46.64 | 38.555 | .855 |
| Data Encryption | 46.00 | 44.000 | .868 |
| Intrusion Detection | 46.07 | 42.687 | .865 |
| Virtual Machine Segregation | 46.79 | 38.027 | .849 |
| Virtual Machine Monitoring | 46.57 | 38.571 | .845 |
| Virtual Machine Migration | 46.64 | 35.632 | .849 |

Table 7 Test for Consistency and Correlation in the First and Second Round Questionnaires.

| | CORRELATIONS | |
| --- | --- | --- |
| | Round One | Round Two |
| Round One Pearson's Correlation | 1 | **0.846** |
| Sig. (2-tailed) | - | **0.001** |
| N | 12 | 12 |
| Round Two Pearson's Correlation | **0.846** | 1 |
| Sig. (2-tailed) | 0.001 | - |
| N | 12 | 12 |

Correlation quantifies the extent to which two quantitative variables, X and Y, "go together." When high values of X are associated with high values of Y, a positive correlation exists. When high values of X are associated with low values of Y, a negative correlation exists (Thanasegaran, 2003), (Malhotra, 2004). In Table 8, the correlation value **r**, is **0.846** which indicates a strong correlation value (usually a correlation value **r,** greater than 0.7 is indicative of a good result). The second component that can be seen in Table 8 is the "Sig" (significance) of the data that was tested. A cut-off mark of **0.05** is used as the yard stick to determine how viable the result of a test is. Hence a significance value is less than 0.05 in our case **0.001**, indicates that the data used for the experiment or test is very good and reliable (Buadu, et al. 2013). After validating the reliability of both round one and round two responses separately, we also established the consistency of both results from the statistical analysis conducted. Hence our result has been proven to be reliable and valid we proceeded to using this information to decide the design of a new secure mobile cloud architecture. In the next section we compare the content value of each security component and use this to categorize the impact factor of these security components.

Table 8 Comparison of the Content Value of the Security Components in Round One and Round Two Questionnaire Response

| Security Component | Content Value 1 | Content Value 2 |
| --- | --- | --- |
| Authentication | 4.80 | 4.64 |
| Authorization | 4.60 | 4.50 |
| Repudiation | 4.27 | 4.14 |
| Network Security | 4.87 | 4.50 |
| Data Security | 4.53 | 4.71 |
| Application Integrity | 4.60 | 4.07 |
| User Management | 4.07 | 3.79 |
| Data Encryption | 4.27 | 4.43 |
| Intrusion Detection | 4.60 | 4.36 |
| Virtual Machine Segregation | 3.73 | 3.64 |
| Virtual Machine Monitoring | 3.73 | 3.86 |
| Virtual Machine Migration | 3.60 | 3.79 |

Based on the outcome/result of the questionnaires from the opinions of the experts displayed/represented graphically in the sections above, we decided to list the important security components into three major categories based on their Content Value (CT). The proposed architecture can be developed in form of an application, in view of which we recognize that the more security components that are embedded in the application, the more the cost of developing the application. On this premise and based on the results, we found that some of the security components are more important than others. Table 9 shows the impact factor/content value scale for classifying each of the security components into three major ranks and is implemented in the proposed architecture shown in Figure 7 and Figure 8 respectively.

Table 9 Categorization of Security Component Based on Content Value

| Content Value | Category |
|---|---|
| 4.50 – 5.00 | Mandatory Component |
| 4.00 – 4.40 | Highly Desirable Component |
| 1.00 – 3.90 | Desirable Component |

# 8 Detailed Description of the Proposed Secure Mobile Cloud Architecture

The architecture contains a group of *cloud security services* which include; User Authorization, User Authentication, Data Security, Data Encryption, Repudiation, Intrusion Detection, User Management, Application Integrity, Network Security, Virtual Machine Monitoring, Virtual Machine Migration and Virtual Machine Segregation. Based on the requirement or situation at hand, each security component is provisioned to carry its functions related to its features or capabilities. For instance the virtual machine monitoring security component is provisioned primarily at the PaaS (Platform as a Service) level where it monitors the virtual machines and triggers the necessary security protocol; if it senses an attack on any of the virtual machines in the pool, an appropriate security component like virtual machine migration kicks in and migrates/isolates the affected virtual machine to another node to wade off the suspected attack

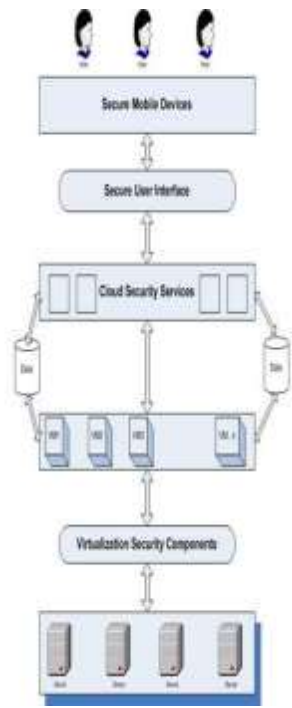

Figure 7 Secure Mobile Cloud Architecture (High-Level) Diagram.

The proposed architecture provides security for the three service layers of the cloud; SaaS (Software as a Service), IaaS (Infrastructure as a Service), PaaS (Platform as a Service) and mobile devices used to access cloud services and resources. That said it can be seen that our proposed architecture is a holistic approach or framework to ensure a secure mobile cloud environment. Relevant security features or components are put in place to mitigate potential security threats. In the IaaS layer, infrastructure security components or measures are put in place for virtualization of related security issues and data center attacks. The security components provisioned for these include; *Virtual Machine Monitoring, Intrusion Detection, Virtual Machine Migration and Virtual Machine Segregation*. The security components provisioned for the *SaaS layer* of the cloud include; *User Authorization, User Authentication, User Management, Repudiation and Application Integrity*. The last but not the least service level is the *PaaS layer* for which security components provisioned to mitigate potential attacks include; *Application Integrity* and virtualization security components (*Virtual Machine Migration, Virtual Machine Monitoring and Virtual Machine Segregation*). The above mentioned security measures are related to specific service levels. These are not the only security measures put in place to ensure a secure mobile cloud. For the management of the entire cloud security, security components that nullify potential security breaches include; *User Authorization*, *User Authentication*, *User Management*, *Repudiation*, *Data Security* and *Data Encryption* and *Intrusion Detection*. We strongly believe that the above security measures are adequate to ensure a secure cloud environment, but there is more to be done to ensure a secure mobile interaction with the cloud. To do this we need to provide security measures for the communication channel which is network security. For the security of the communication channel, we use *Data Security*, *Data Encryption* and *Network Security*; using a secure routing protocol. Furthermore it is also pertinent that the mobile device which will be used to access cloud services and resources is secure. For this, we also will be using some of the security components that are used in providing a secure cloud environment which include; *Application Integrity* (applications installed on the mobile device), *User Access Authentication and Authorization*, *Repudiation*, *Data Encryption* and *Data Protection*.

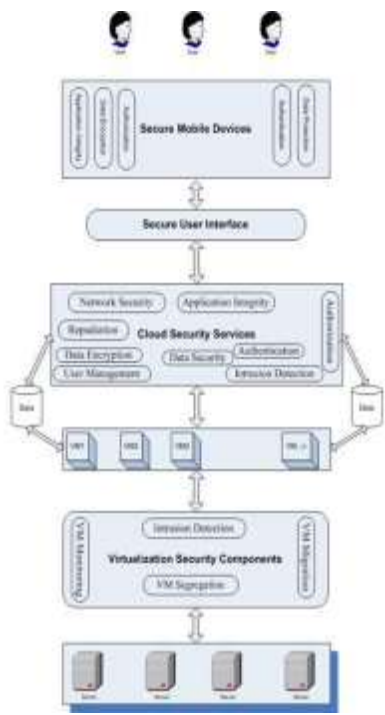

Figure 8 Secure Mobile Cloud Architecture (Low-Level) Diagram.

## 9. Conclusion

Through our interaction with people in various works of life; academia, information technology investors, students, lecturers and tech savvy people, we found out that there is much hype and controversy surrounding both mobile cloud computing and cloud computing. Yes it is true that there exist real issues that call for concern in connection with this new breed of technology but some of these concerns are mere rumors without any basis of justification. One of the major worries relating to this important subject of interest (mobile cloud computing) is that of security (KPMG, 2010), (Zissis & Lekkas, 2012) which this research has focused on. In conducting this research we have made an effort to demystify the security challenges that plague this evolving technology and have gone a step further to proffer possible mitigation plans or solutions to debunk these security concerns. Through a thorough review of relevant literature, we have been able to get facts and figures regarding the potential threats of investing in mobile cloud computing. We did not end our fact finding drive on just information gathered from academic research or materials, we went a step further to contact established academicians in the field of mobile cloud computing and cloud computing to validate our earlier findings. Armed with information from trusted parties we were able to establish what some of the real concerns are with respect to adopting mobile cloud computing. With a clear mandate in view, we continued with researching and sending out questionnaires to identify some of the ways through which the established concerns can be resolved so as to present a possible solution to restoring trust in mobile cloud computing, considering the immense benefits of mobile cloud computing and cloud computing; a foremost research group has tipped cloud computing which is the bed rock for mobile cloud computing as the first amongst the top ten most important technologies with a better prospect with time (Gartner Inc., 2011). In order to present the potential security threats of both mobile cloud computing and cloud computing in a concise, manageable and easy to relate to way, we have classified the security threats into three major categories:

I. Mobile device threats
II. Threats to the cloud (Cloud computing)
III. Network threats (Communication channel)

We have outlined mitigations plans and security components to address in details each category of security issues that are concerned with mobile cloud computing and we have designed a framework/architecture with security components to counter attacks on the cloud or mobile devices used to access cloud services or resources.

Acknowledgments

The authors thank Dr. Arash Habibi Lashkari and Mr. Chris Roadknight for their support in this project.

IJCSI International Journal of Computer Science Issues, Volume 12, Issue 2, March 2015
ISSN (Print): 1694-0814 | ISSN (Online): 1694-0784
www.IJCSI.org